\documentclass[runningheads]{llncs}
\usepackage[T1]{fontenc}
\usepackage{hyperref} 
\usepackage{tabularx} 
\newcolumntype{C}{>{\centering\arraybackslash}X}

\usepackage{float}
\usepackage{xcolor}
\usepackage{graphicx}
\usepackage{subcaption}
\begin{document}
%

\title{Architectural Trade-offs in the Energy-Efficient Era: A Comparative Study of power-capping NVIDIA H100 and H200}
\titlerunning{Architectural Trade-offs in the Energy-Efficient Era}
%
%
%
\author{Aditya Ujeniya\inst{1}\orcidID{0009-0000-1670-4131} \and
Jan Eitzinger\inst{1}\orcidID{0009-0000-3350-3841} \and \\
Georg Hager\inst{1}\orcidID{0000-0002-8723-2781} \and
Gerhard Wellein\inst{1}\orcidID{0000-0001-7371-3026}}
\authorrunning{A. Ujeniya et al.}
%
\institute{\textsuperscript{1} Erlangen National High Performance Computing Center (NHR@FAU)\\Friedrich-Alexander University of Erlangen–Nuremberg, 91058 Erlangen, Germany 
\email{\{aditya.ujeniya, jan.eitzinger, georg.hager, gerhard.wellein\}@fau.de}}
\maketitle              
\begin{abstract}
Modern NVIDIA GPUs like the H100 (HBM2e) and H200 (HBM3e) share similar compute characteristics but differ significantly in memory interface technology and bandwidth. By isolating memory bandwidth as a key variable, the power distribution between the memory and Streaming Multiprocessors (SM) changes notably between the two architectures. In the era of energy-efficient computing, analyzing how these hardware characteristics impact performance per watt is critical. This study investigates how the H100 and H200 manage memory power consumption at various power-cap levels. By a regression analysis, we study the memory power limit and uncover outliers consuming more memory power. To evaluate efficiency, we employ compute-bound (DGEMM) and memory-bound (TheBandwidthBenchmark) workloads, representing the two extremes of the Roof\-line model. Our observations indicate that across varying power caps, the H100 remains the slightly better choice for strictly compute-bound workloads, whereas the H200 demonstrates superior efficiency for memory-bound applications. We then perform similar experiments on an AI workload, a Vision Transformer model, and try to apply the insights from our micro-benchmark analysis.

\end{abstract}
\section{Introduction}

Power consumption in High-Performance Computing (HPC) and Artificial Intelligence (AI) has become a primary bottleneck for data centers. To manage these power demands, administrators frequently use deterministic hardware power-capping to limit GPU power draw. However, predicting performance under these power caps is difficult because modern GPUs employ complex Dynamic Voltage and Frequency Scaling (DVFS) algorithms to arbitrate power internally. Power consumption generally varies with arithmetic intensity. Synthetic benchmarks representing instruction throughput- and memory bandwidth-limited scenarios can systematically probe Roof\-line model corner cases under different power caps.

This study investigates the performance and energy efficiency of the NVIDIA H100 and H200 GPUs under different power caps. Although both GPUs are built on the Hopper architecture and share similar compute capabilities, they differ significantly in memory bandwidth (HBM2e vs. HBM3e). By evaluating these GPUs using compute-bound (DGEMM) and memory-bound (TheBandwidthBenchmark) workloads, we isolate the impact of memory bandwidth on power distribution between the Streaming Multiprocessors (SM) and the memory subsystem. Our goal is to provide an empirical analysis of how varying power limits affect performance, frequency stability, and overall efficiency across these two architectures. 

We start by introducing the GPU configuration in Section \ref{sec:3a}. We can see that the peak performance numbers of the NVIDIA H100 and H200 are identical. The only difference is the amount of memory and the memory bandwidth. Sections \ref{sec:3b} and \ref{sec:3c} also describe briefly the synthetic benchmarks used to perform the regression analysis. They reflect the two different corner cases in the Roof\-line model: DGEMM for the compute-bound regime and Schönauer Triad for the  memory-bound regime. To apply our understandings from the micro-benchmark analysis, we perform similar power capping experiments on an AI workload, a Vision Transformer (ViT) model. Section \ref{sec:3d} details the parameters used for the ViT benchmark. Section \ref{sec:3e} covers the methodology for performing measurements, which is the basis for collecting different metrics and discussing valid assumptions for this study.

Section \ref{sec:4} explores the GPU power and memory power draw, frequency throttling effects, and their impact on the benchmark score under different power-cap settings. The regression of the benchmarks also allows us to pinpoint outliers among specimens as well as understand the maximum memory power draw, which is not officially documented by NVIDIA. Section \ref{sec:5} presents an energy efficiency analysis in terms of performance per Watt, comparing the power breakdown side by side for both the GPUs.
 
\section{Related Work}

Prior work on power-capping GPUs shows that enforcing power limits saves energy more reliably than reducing clock speeds, as demonstrated on NVIDIA Volta hardware \cite{Patki}. With the rise of large-scale AI workloads, recent research has focused on energy efficiency under strict power budgets \cite{Yuan}, finding that maximum throughput does not always minimize energy consumption. In heterogeneous CPU-GPU systems, proper workload balance is critical, as CPU bottlenecks can undermine power-capping effectiveness \cite{Oscar,Adam}.

GROMACS simulations on A100 and L4 GPUs show that performance is rather independent of power caps, degrading only under tight limits \cite{Ayesha}. Together, these findings confirm that high-end GPUs can sustain near-peak application performance at substantially reduced power levels.

While older GPU generations are well studied, the newer Hopper architecture remains underexplored. Recent work by Mayr et al. \cite{Martin} compared H100 and H200 GPUs under strict power caps, but their evaluation focused exclusively on AI benchmarks. Consequently, it remains unclear how the differing memory technologies of these GPUs (HBM2e vs.\ HBM3e) affect energy efficiency across other types of workloads. Furthermore, previous work did not analyze internal hardware behaviors, such as memory power draw or SM frequency throttling, under power constraints. This study addresses these gaps by isolating memory bandwidth to assess its direct impact on performance-per-watt. We provide a detailed analysis of memory power consumption and throttling mechanics across a wide range of power cap settings.

\section{Testbed and Benchmarks}
In this section, we outline the hardware characteristics of the NVIDIA GPUs and detail our benchmarking methodology. To ensure statistical reliability, we execute 50 complete benchmark runs for every power-cap setting on each individual GPU. Across our testing environment of four nodes (each equipped with four GPUs), this yields 800 discrete measured data points per benchmark for each specific power cap. The specific performance and hardware metrics captured during these iterations are detailed in Section \ref{sec:3e}.

\subsection{System Characteristics of the Helma cluster}
\label{sec:3a}
As shown in Table~\ref{tab1}, the two GPU variants are very similar except for the memory characteristics. The H200 has clearly better memory technology: HBM3e with more stacks and higher memory frequency. More information on the node configuration can be found in the vendor documentation for the compute node.\footnote{https://lenovopress.lenovo.com/lp1613-thinksystem-sd665-n-v3-server}

\setlength{\tabcolsep}{10pt} 
\renewcommand{\arraystretch}{1.2} 
\begin{table}[tbp]
\centering
\caption{Characteristics of the NVIDIA H100 and H200 used for benchmarks}\label{tab1}
\begin{tabularx}{\textwidth}{| l | C | C |} \hline
\textbf{Metric} & \textbf{H100} & \textbf{H200} \\ \hline
Power Draw & 700\,W & 700\,W\\ 
Memory Capacity & 94 GiB & 144 GiB \\ 
Memory technology & HBM2e & HBM3e \\ 
Base SM Frequency & 1665 MHz & 1665 MHz \\ 
Boost SM Frequency & 1980 MHz & 1980 MHz \\ 
Memory Frequency & 1593 MHz & 3201 MHz \\ 
TF64 Peak & 67 TFlop/s & 67 TFlop/s \\ 
Mem BW & 2.41 TB/s & 4.89 TB/s \\  \hline
Nodes used & \multicolumn{2}{|c|}{4} \\ 
GPUs per node  & \multicolumn{2}{|c|}{4} \\ 
Performance Mode & \multicolumn{2}{|c|}{P0} \\
Software Stack & \multicolumn{2}{|c|}{NVHPC SKD 24.11 and CUDA 12.9} \\ \hline
\end{tabularx}
\end{table}

\subsection{General Matrix Multiplication (GEMM)}
\label{sec:3b}
The NVIDIA Hopper architecture (H100 and H200) uses fourth-generation Tensor Cores, which provide native FP64 support to drastically speed up double-precision math operations rather than relying purely on standard CUDA cores. We used the \texttt{cuBLAS}\footnote{https://developer.nvidia.com/cublas} library to perform double-precision general matrix-matrix multiplications (DGEMM), $C = \alpha AB + \beta C$. We use DGEMM to mimic compute-bound codes. For the H100 and H200 DGEMM benchmarks, we used square matrices of size 32,768 ($2^{15}$, a multiple of 32). This problem size is large enough to not fit into the cache, leading to significant (but not saturated) memory bandwidth. Furthermore, while outside the strict scope of this study, it is worth highlighting that increasing the DGEMM problem size on the H100 makes the Tensor Cores increasingly memory-bound at different power cap levels. Consequently, we selected this specific matrix size to maximize workload size just before hitting that severe memory bottleneck. The optimal thread-block size is automatically determined by the cuBLAS library.

\subsection{TheBandwidthBenchmark (TBB)}
\label{sec:3c}
\begin{figure}[tbp]
    \centering
    \includegraphics[width=1\textwidth]{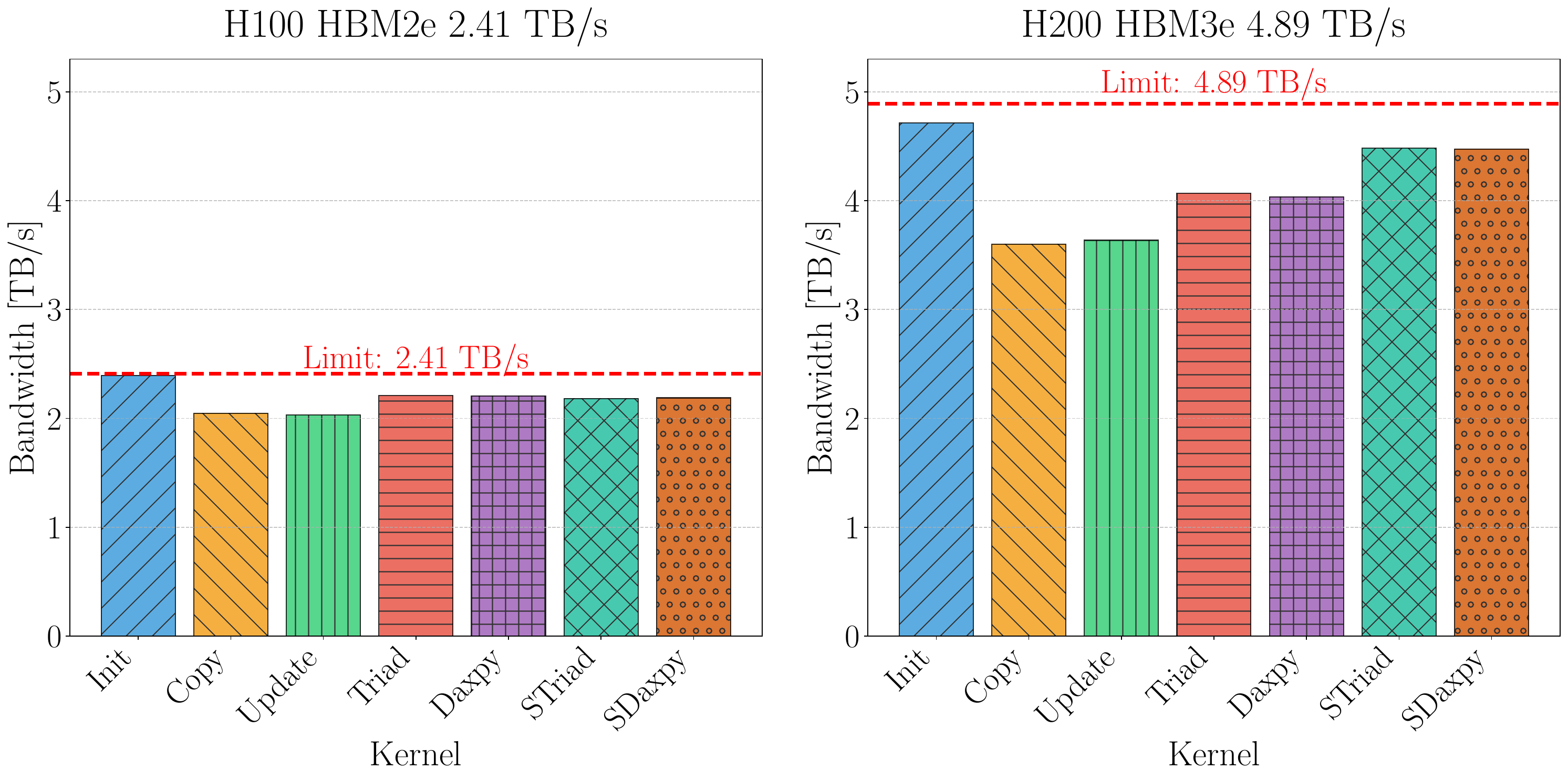}
    \caption{Effective bandwidth for different kernels on NVIDIA H100 and H200}
    \label{fig:effbw}
\end{figure}

TheBandwidthBenchmark\footnote{https://github.com/HPC-Dwarfs/TheBandwidthBenchmark} is a flexible benchmarking suite designed to measure the sustained effective main memory bandwidth on both CPU and GPU architectures. It is inspired by the McCalpin STREAM benchmark but adds more kernels to evaluate memory performance across various data access patterns (note that \texttt{s} is a scalar value):\\[1mm]
\begin{tabular}{@{}l@{\hspace{0.5em}}l@{\hspace{1em}}l@{\hspace{0.5em}}l@{}}
\textbf{Init}: & \texttt{a[i] = s} & \textbf{Daxpy}: & \texttt{a[i] = a[i] + b[i] * s} \\
\textbf{Copy}: & \texttt{a[i] = b[i]} & \textbf{STriad}: & \texttt{a[i] = b[i] + c[i] * d[i]} \\
\textbf{Update}: & \texttt{a[i] = a[i] * s} & \textbf{SDaxpy}: & \texttt{a[i] = a[i] + b[i] * c[i]} \\
\textbf{Triad}: & \texttt{a[i] = b[i] + c[i] * s} & & \\
\end{tabular}\\[1mm]
Figure~\ref{fig:effbw} shows the effective bandwidths for different kernels on NVIDIA H100 and H200.
For our purposes, we will focus on the double-precision Schönauer Triad (STriad) kernel performance. 
For all STriad benchmarks on H100 and H200, we use a total data set size of 80 GB. Although a problem size in excess of 4~GB would have sufficed to make it memory bound, we choose a problem size that takes the better part of the memory on each GPU. The optimal thread-block size is automatically determined by TBB.

\subsection{Vision Transformer (ViT) model}
\label{sec:3d}
For our AI workload analysis, we evaluate the Vision Transformer (ViT-L/16) \footnote{https://github.com/RRZE-HPC/hpc-ai-perf-bench/tree/main/computer\_vision} architecture as a representative attention-heavy, transformer-based model. Following the standardized, power-aware benchmarking suite proposed by Mayr et al. \cite{Martin}, the framework isolates steady-state hardware compute limits by measuring raw application-level throughput, specifically captured in images per second. To avoid skewing results with system overhead, the first epoch is discarded as a warm-up phase to eliminate initialization transients and Just-In-Time (JIT) compilation effects.

To achieve exact reproducibility and ensure proper utilization of the underlying compute hardware, execution parameters must be precisely controlled. The benchmark uses a large, memory-saturating batch size of 256 designed to bypass caching and maximize utilization of the hardware's Tensor Cores. We used a uniform software stack of CUDA 12.9 and PyTorch 2.x and the model runs under TF32 mixed-precision across the H100 and H200 platforms. The ViT benchmark is a multi-GPU benchmark which benefits from multi-GPU execution. To effectively compare against the micro-benchmarks, which are single GPU only, we perform the experiments for the ViT benchmark on single GPU only. The performance and the efficiency may look different if ViT was executed as a multi-GPU benchmark due to inter-GPU communication. The ViT benchmark is neither completely memory bound like the Schönauer Triad nor is it compute bound like DGEMM; it lies in between,  with slightly more dependency on memory bandwidth.

\subsection{Methodology}
\label{sec:3e}

We executed the micro-benchmarks independently across all GPUs on four nodes, i.e.,  identical binaries were submitted on 16 GPUs at once with different power caps. For the ViT benchmark, we perform regression on 1 GPU instead of 16 GPUs due to limited hardware availability. For each execution, we capture the benchmark throughput (in TFlop/s or TB/s or samples/sec), along with average metrics for total power draw (W), memory power draw (W), SM frequency (MHz), and memory frequency (MHz). According to the NVIDIA NVML documentation for \texttt{nvmlDeviceGetPowerUsage}~\cite{nvidia_nvml}, the power draw reported is for the whole GPU including associated circuitry like memory, etc. Hence, we assume that the overall power consumption reported $W_\mathrm{Total}$ is additive:
\begin{equation}
W_{\mathrm{Total}} = W_{\mathrm{GPU}} + W_{\mathrm{Memory}}
\label{eq:power_breakdown}
\end{equation}
Since both of our micro-benchmarks run for at least 10 minutes at full Thermal Design Power (TDP) for a single run, both $W_{\mathrm{Total}}$ and $W_{\mathrm{Memory}}$ are recorded using \texttt{nvidia-smi -q} with a 10-second polling interval. Similar metrics for the ViT benchmark are collected using a 100\,ms polling interval since the workload is not constant over time. $W_{\mathrm{GPU}}$ is the power draw combining the SM units, the L2 cache, and the memory controller. At the end of each benchmark run, the collected values are time-averaged to generate one consolidated data point per benchmark run. Because there is no inter-GPU communication during these runs, communication-related power draw is excluded from our analysis.

We evaluated both the NVIDIA H100 and H200 across a power-capping range from 200\,W to their maximum TDP of 700\,W in increments of 100\,W. Across all tested GPU power limits, the memory frequency remained strictly static: 1593~MHz for the H100 and 3201~MHz for the H200.

All experiments use random data initialization to maximize power draw. Randomly initialized data causes more dynamic power due to more frequent transistor switching~\cite{Anantha,Najm}; we experimented with different array initialization techniques like constant and random data and found up to 100\,W difference in power draw. This is in line with similar studies on recent CPUs and GPUs~\cite{schoene2019,Oscar}, but it is not within the scope of this study to go into detail of this behavior. More data on this topic is available online.\footnote{https://blogs.fau.de/adityauj/2025/11/15/power-frequency-characteristics/}.

\section{Power-capping analysis for different benchmarks}
\label{sec:4}

For each benchmark, we show power vs.\ performance across different power-cap settings. Then we analyze the observed behavior with split violin plots for SM frequency and memory power draw. All charts include regression data (repeated end-to-end measurements and GPU specimens). 
There is interesting spatial (cross-specimens) behavior which is not covered since extensive regression analysis is not in the scope of this study. However it is worthwhile to point out that we aggregate metric results from multiple distinct GPUs, so the plots exhibit notable cross-specimen spatial variance at different power caps. The variance of the metrics within one GPU across different power caps and across time is smaller than 5\%. We have excluded baseline power draw, such as CPU power or other components beyond the GPUs. Note that if we had enforced a power limit on the entire node instead of just the GPU, the results would look different.

\subsection{DGEMM}

\begin{figure}[tb]
    \centering
    \begin{subfigure}[b]{0.49\textwidth}
        \includegraphics[width=\textwidth]{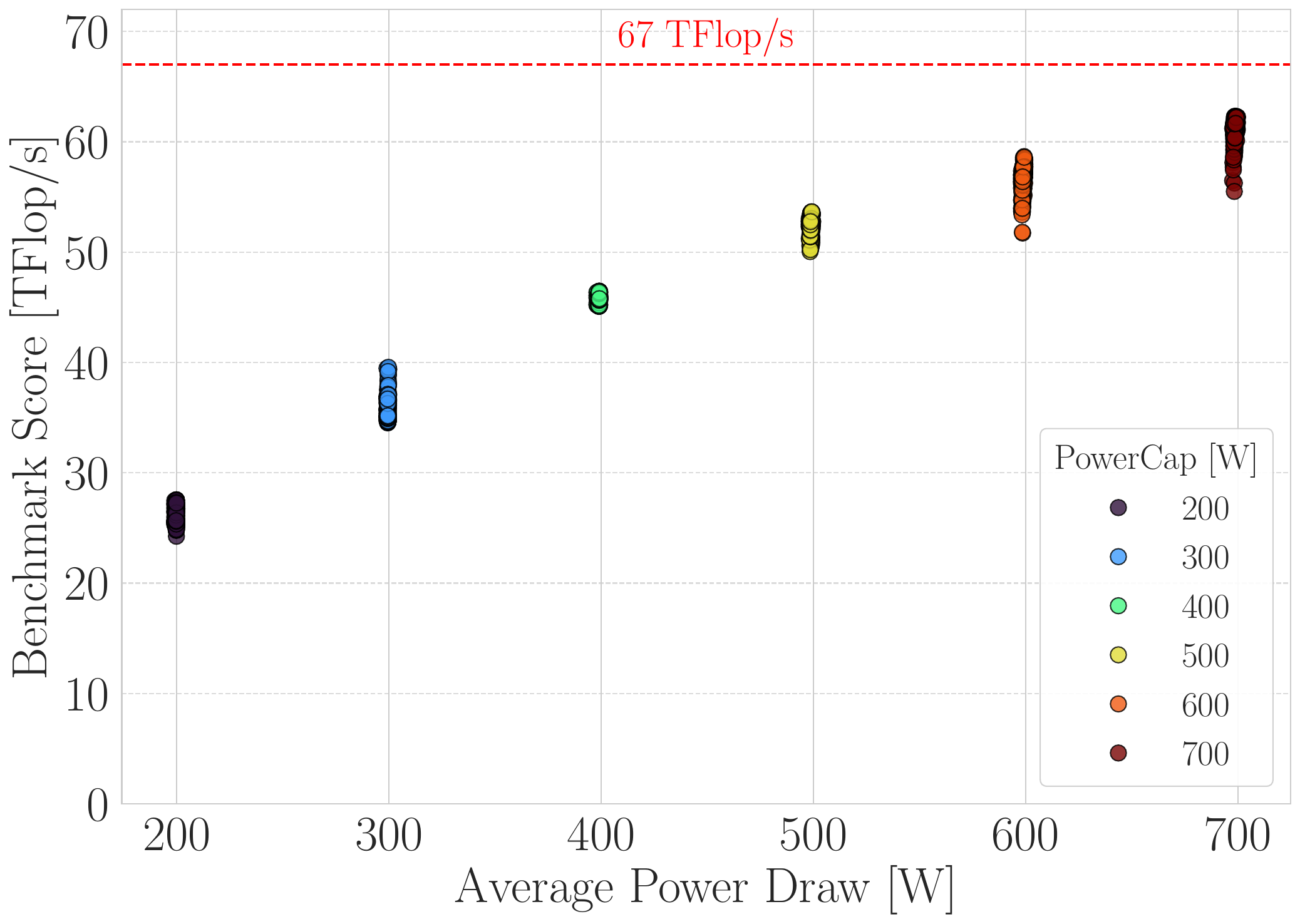}
        \caption{NVIDIA H100}
        \label{fig:zplot_h100}
    \end{subfigure}
    \hfill 
    \begin{subfigure}[b]{0.49\textwidth}
        \includegraphics[width=\textwidth]{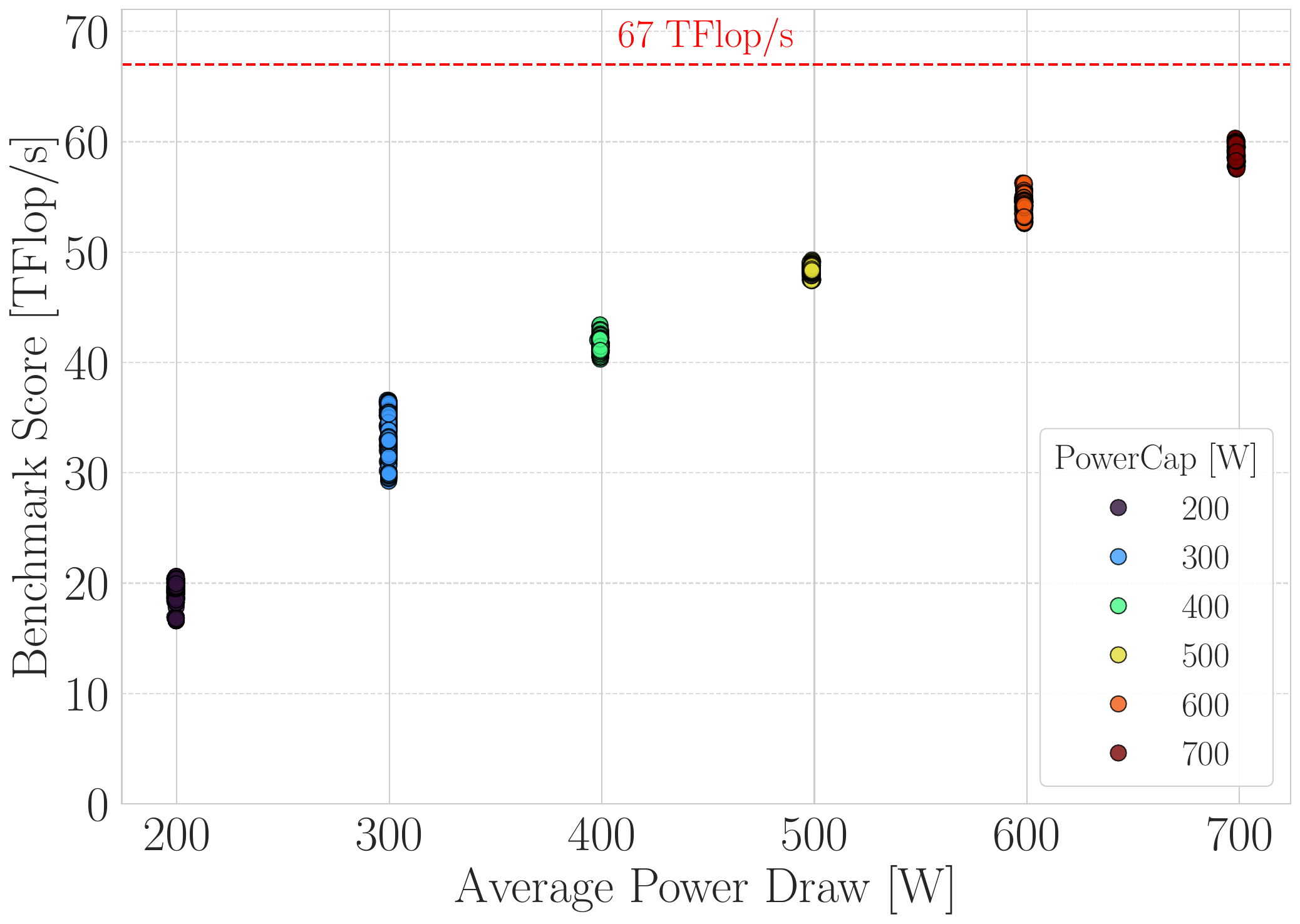}
        \caption{NVIDIA H200}
        \label{fig:zplot_h200}
    \end{subfigure}
    \caption{DGEMM: relationship between performance throughput and average power draw across different power-cap settings.}
    \label{fig:zplot_dgemm}
\end{figure}
\begin{figure}[tb]
    \centering
    \begin{subfigure}[b]{0.49\textwidth}
        \includegraphics[width=\textwidth]{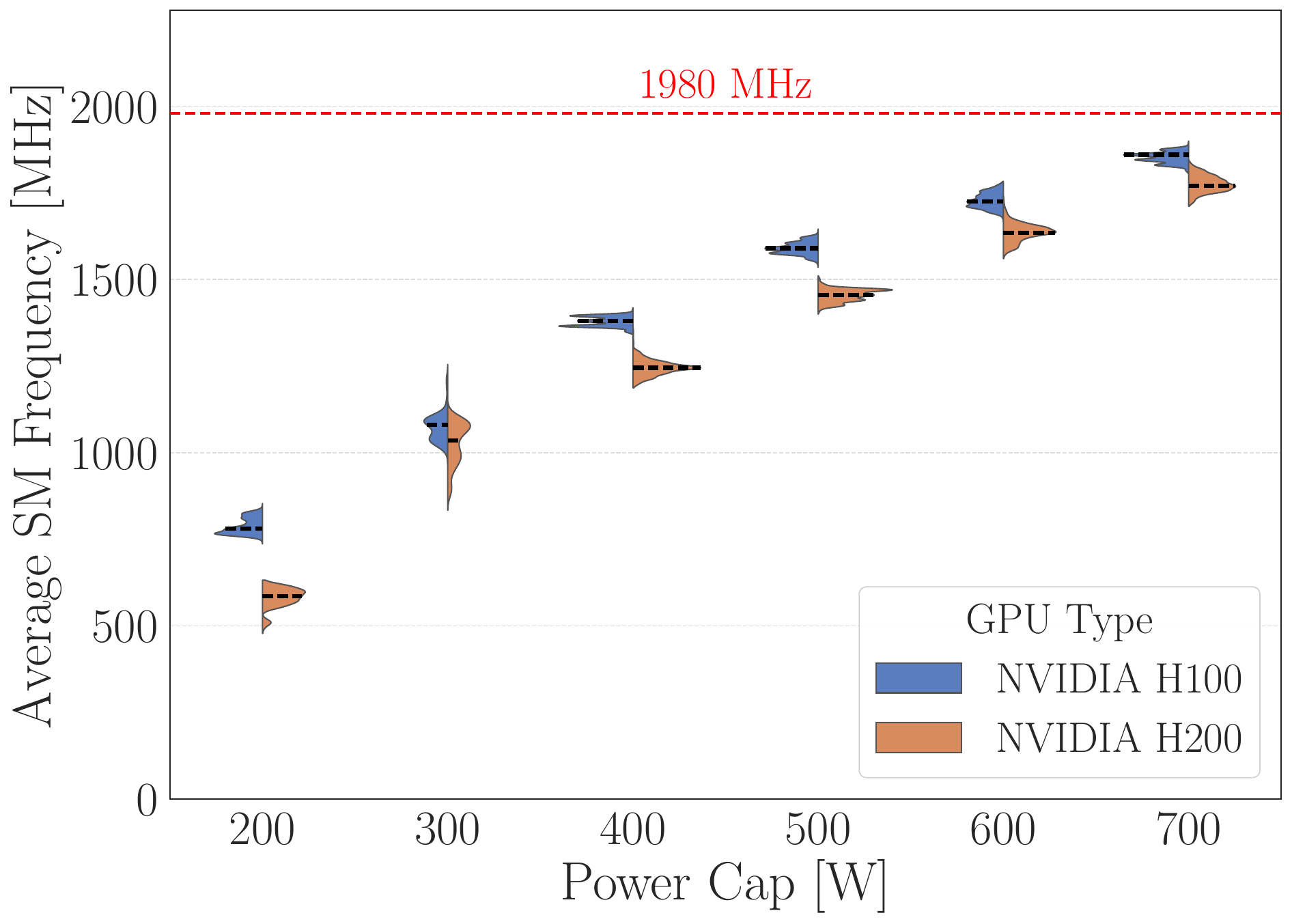}
        \caption{Average SM Frequency}
        \label{fig:SV_DGEMM_Freq}
    \end{subfigure}
    \hfill 
    \begin{subfigure}[b]{0.49\textwidth}
        \includegraphics[width=\textwidth]{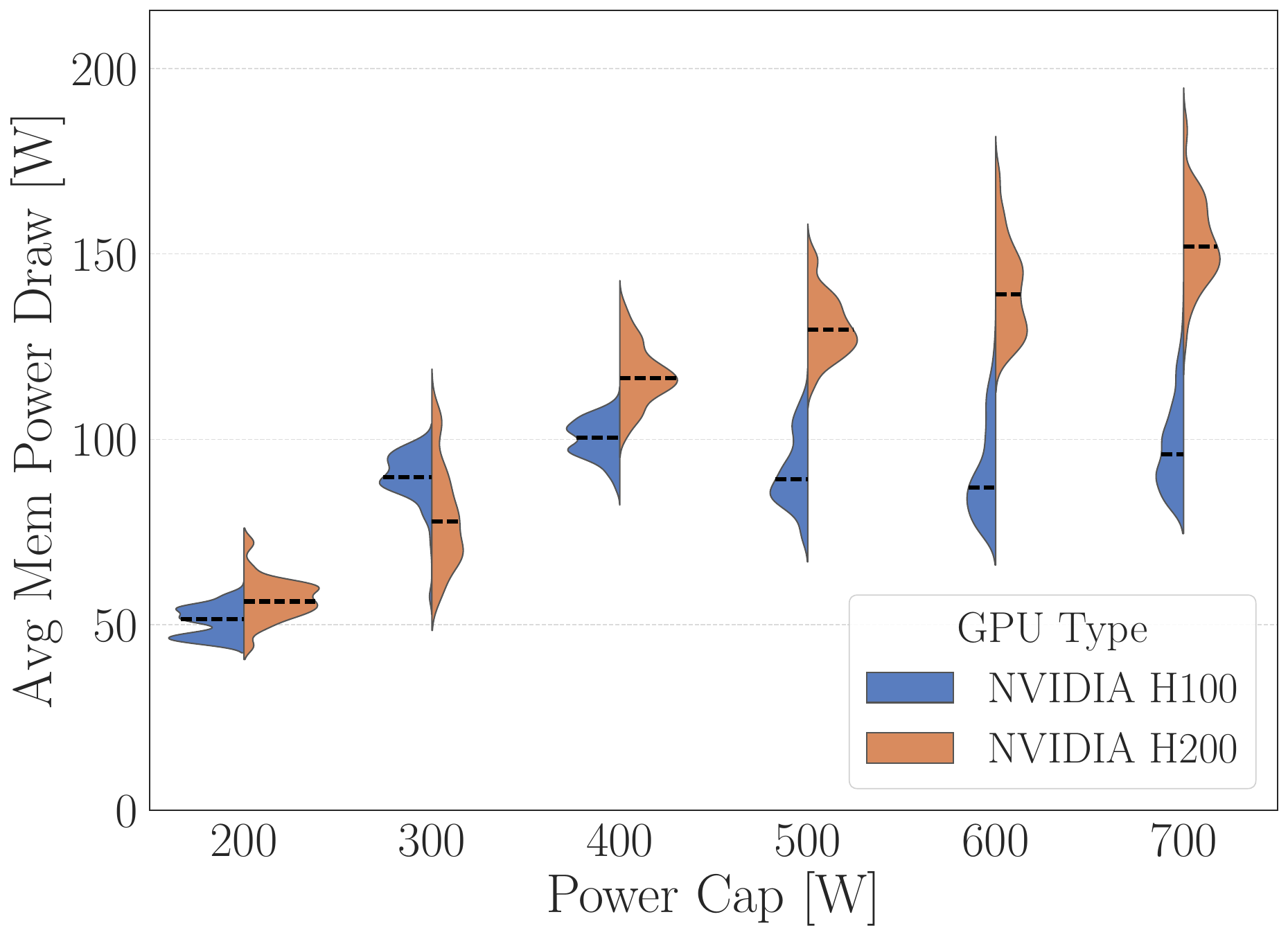}
        \caption{Average Memory Power Draw} 
        \label{fig:SV_DGEMM_MemP}
    \end{subfigure}
    \caption{DGEMM: split violin plots detailing average SM frequency and memory power draw.}
    \label{fig:SV_DGEMM}
\end{figure}

Figure~\ref{fig:zplot_dgemm} reveals that, while the DGEMM benchmark consistently saturates the power cap (showing minimal variance on the power draw axis), its performance fluctuates significantly. The most stable performance zones occur at 400\,W for the H100 and at 500\,W for the H200. Scaling the total power cap from 200\,W to 400\,W yields substantial performance gains, whereas the 500\,W to 700\,W range exhibits diminishing returns, offering only a 10\% performance increase at each power-cap step. Notably, the H100 slightly outperforms the H200 at identical power limits. This advantage for the H100 stems from the H200's higher average memory power draw (see Figure~\ref{fig:SV_DGEMM_MemP}), which reduces the available power budget for the SMs. Consequently, the H200 experiences more aggressive SM frequency throttling, which disproportionately degrades the performance of compute-bound workloads like DGEMM.

Figure~\ref{fig:SV_DGEMM} further illustrates this inverse relationship between memory power draw and SM frequency. This inverse relation is observed when GPUs are throttling. Because the H200 operates at a higher memory frequency, it inherently draws more memory power. Interestingly, while the H100's average memory power draw plateaus between 100\,W and 110\,W, the H200's memory power draw continues to scale concurrently with the total power cap. A notable exception occurs precisely at the 300\,W power-cap setting, where both GPUs exhibit nearly identical memory power consumption and average SM frequencies.

\subsection{Schönauer Triad}

\begin{figure}[tbp]
    \centering
    \begin{subfigure}[b]{0.49\textwidth}
        \includegraphics[width=\textwidth]{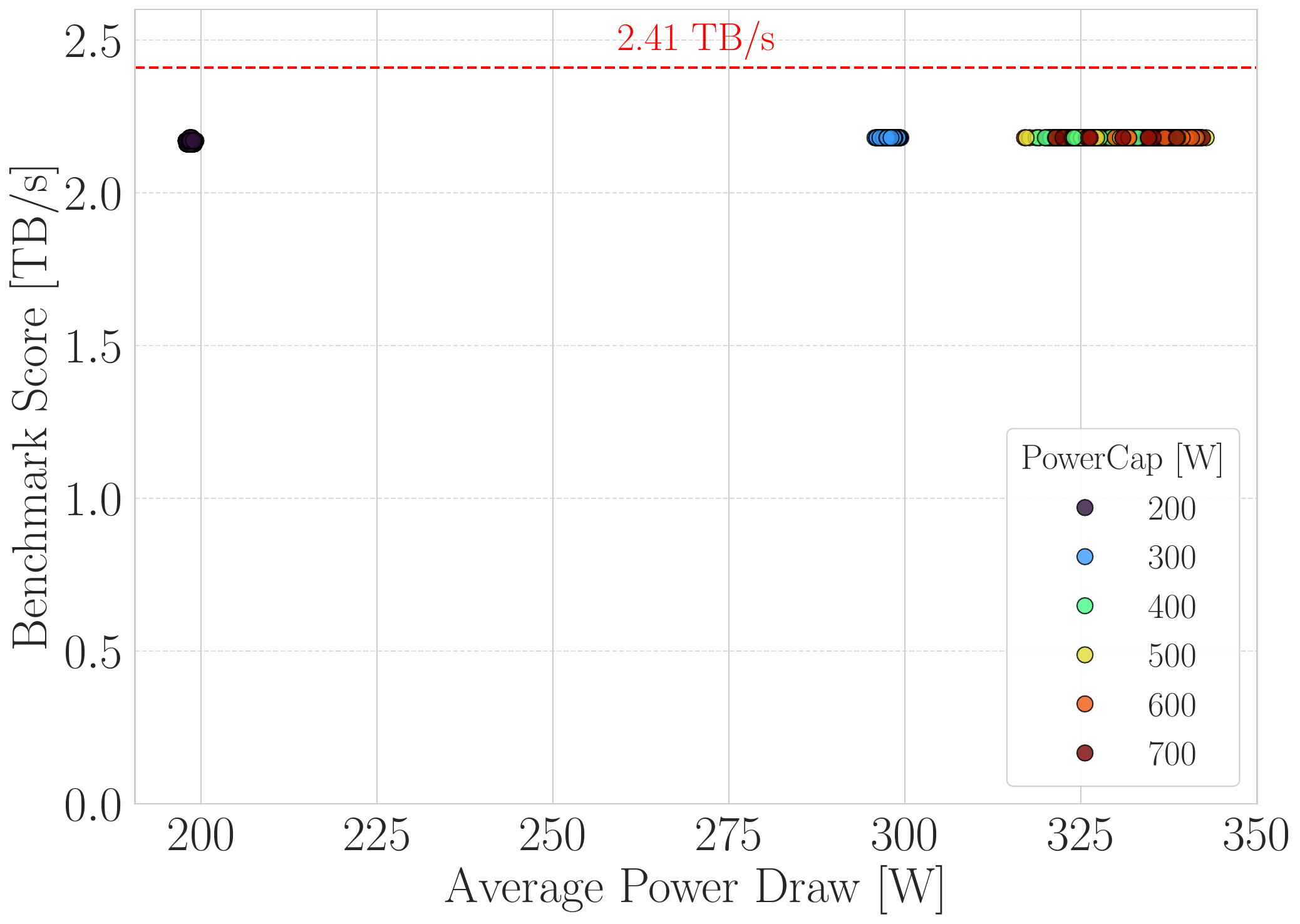}
        \caption{NVIDIA H100}
        \label{fig:zplot_striad_h100}
    \end{subfigure}
    \hfill 
    \begin{subfigure}[b]{0.49\textwidth}
        \includegraphics[width=\textwidth]{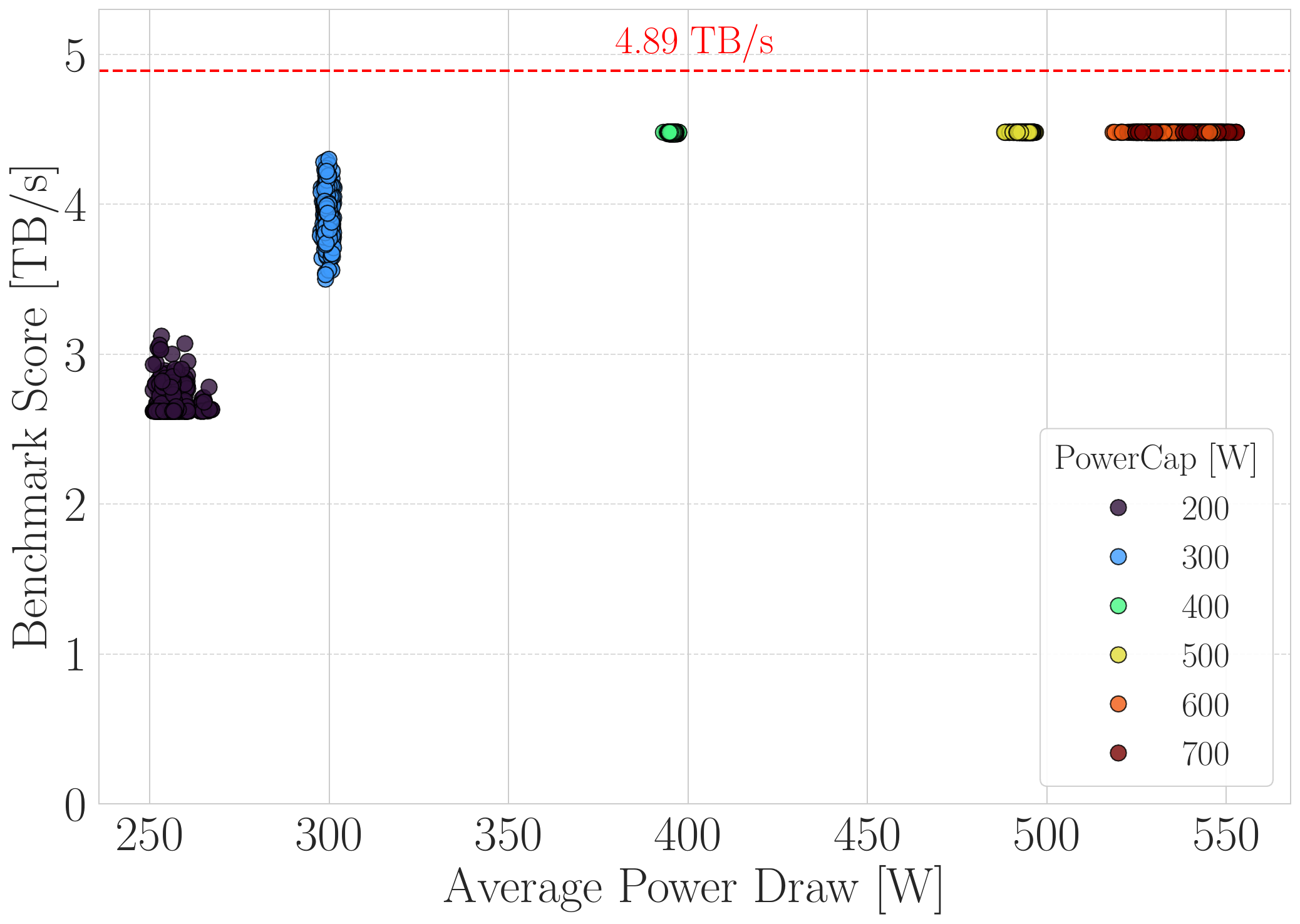}
        \caption{NVIDIA H200}
        \label{fig:zplot_striad_h200}
    \end{subfigure}
    \caption{Relationship between performance throughput and power draw across different power-cap settings for the Schönauer Triad benchmark.}
    \label{fig:zplot_striad}
\end{figure}
In contrast to compute-bound workloads, memory-bound loops such as STriad do not fully saturate the GPU's TDP. To achieve peak effective memory bandwidth, the NVIDIA H100 requires only around 350\,W. Conversely, the H200 demands a significantly higher average power draw of at least 550\,W to sustain its larger memory bandwidth. Notably, the H100 maintains peak effective bandwidth even under a strict 200\,W power cap, whereas the H200 exhibits bandwidth degradation at limits of 300\,W and below.

\begin{figure}[tbp]
    \centering
    \begin{subfigure}[b]{0.49\textwidth}
        \includegraphics[width=\textwidth]{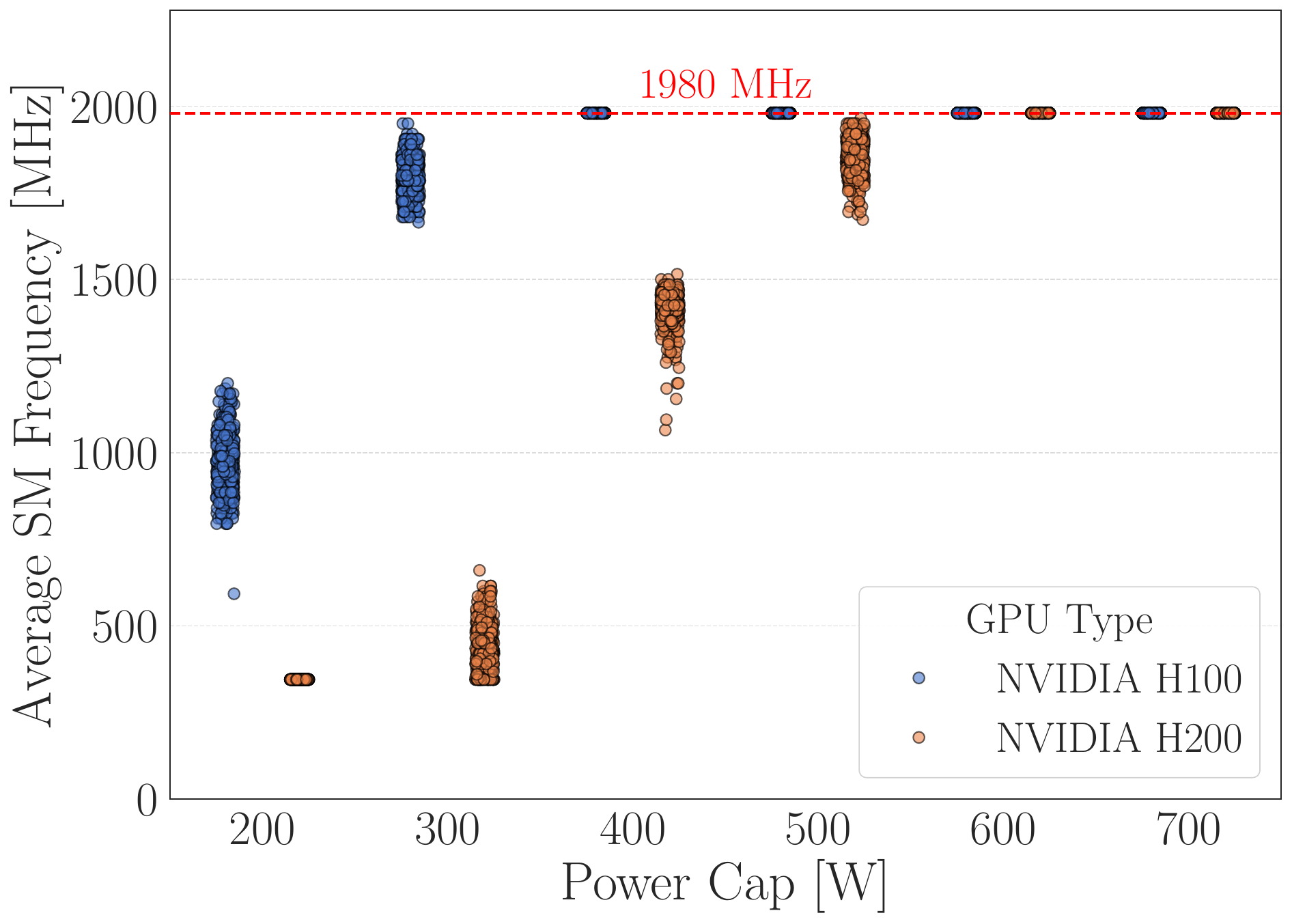}
        \caption{Average SM Frequency}
        \label{fig:sv_striad_freq}
    \end{subfigure}
    \hfill 
    \begin{subfigure}[b]{0.49\textwidth}
        \includegraphics[width=\textwidth]{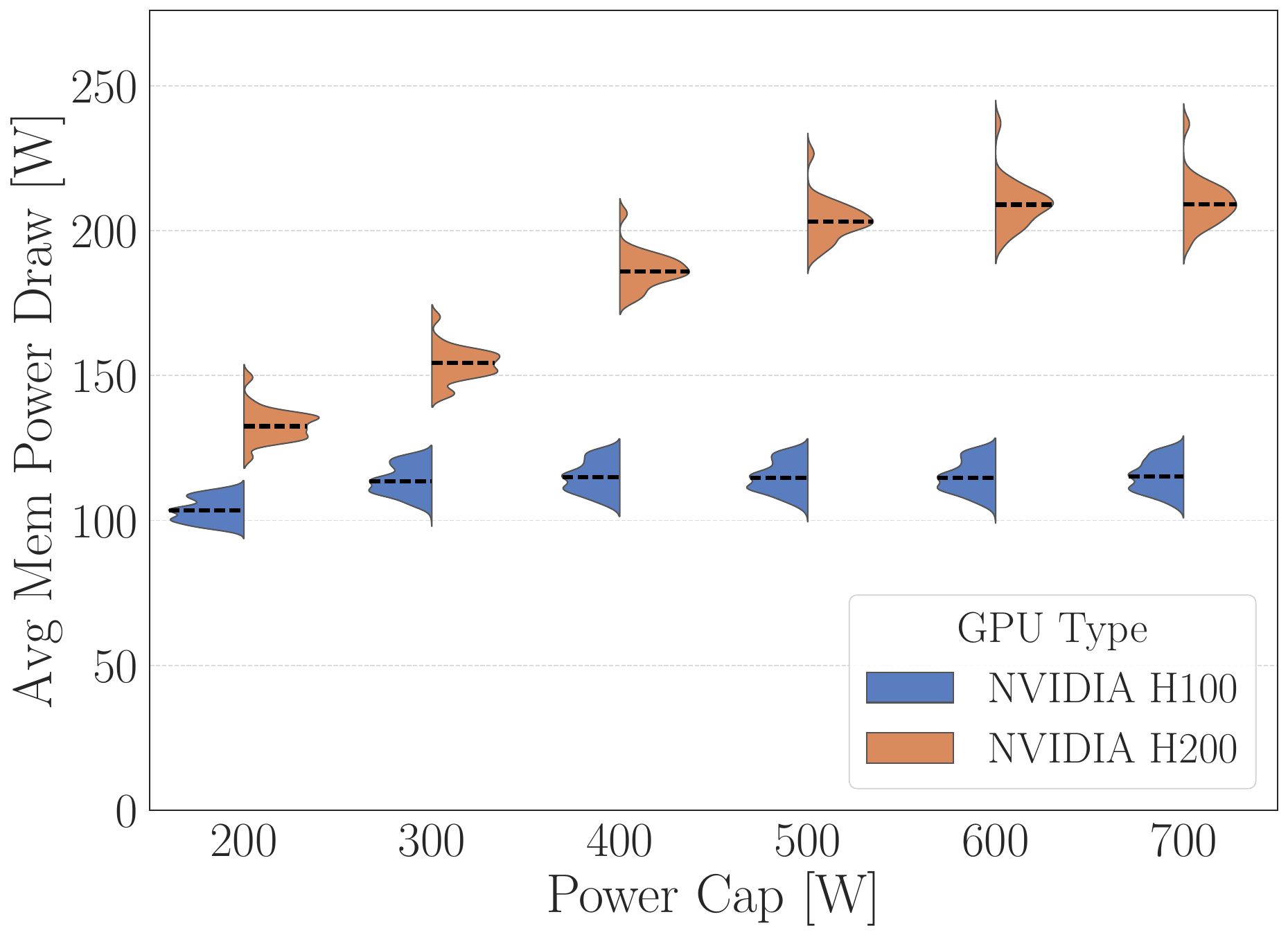}
        \caption{Average Memory Power Draw}
        \label{fig:sv_striad_mempower}
    \end{subfigure}
    \caption{Schönauer Triad performance metrics comparing the NVIDIA H100 (left halves) and H200 (right halves) via split scattered and violin plots.}
    \label{fig:sv_striad}
\end{figure}
Furthermore, Figure~\ref{fig:zplot_striad_h200} reveals anomalous behavior for the H200 at the 200\,W setting: the actual average power draw hovers around 250\,W. Unlike compute-bound scenarios (Figure~\ref{fig:zplot_h200}), the H200 fails to strictly enforce the 200\,W limit during memory-bound execution, a phenomenon further explored in Section~\ref{STriad_efficiency}.

Notably, SM frequency throttling reduces effective memory bandwidth as the GPU approaches its base clock limit. As shown in Figure~\ref{fig:sv_striad_freq}, the H200 experiences severe throttling down to its 345~MHz base operating frequency, a state the H100 never reaches. This aggressive throttling in the H200 is driven by its higher memory power requirements. Figure~\ref{fig:sv_striad_mempower} illustrates that the H100's average memory power draw saturates at approximately 125 W, while the H200 saturates much higher at roughly 220\,W. 

Examining regressions also reveals hardware outliers within the power metrics. For example, Figure~\ref{fig:sv_striad_mempower} shows a distinct secondary distribution near 10\% higher memory power for the H200 across all power caps, indicating a specific GPU unit consistently drawing more memory power than other GPUs. This hardware-level variance (second distribution created by one GPU consistently drawing more power) is also visible in the compute-bound results in Figure~\ref{fig:SV_DGEMM_MemP}.

\subsection{ViT}

\begin{figure}[tbp]
    \centering
    \begin{subfigure}[b]{0.49\textwidth}
        \includegraphics[width=\textwidth]{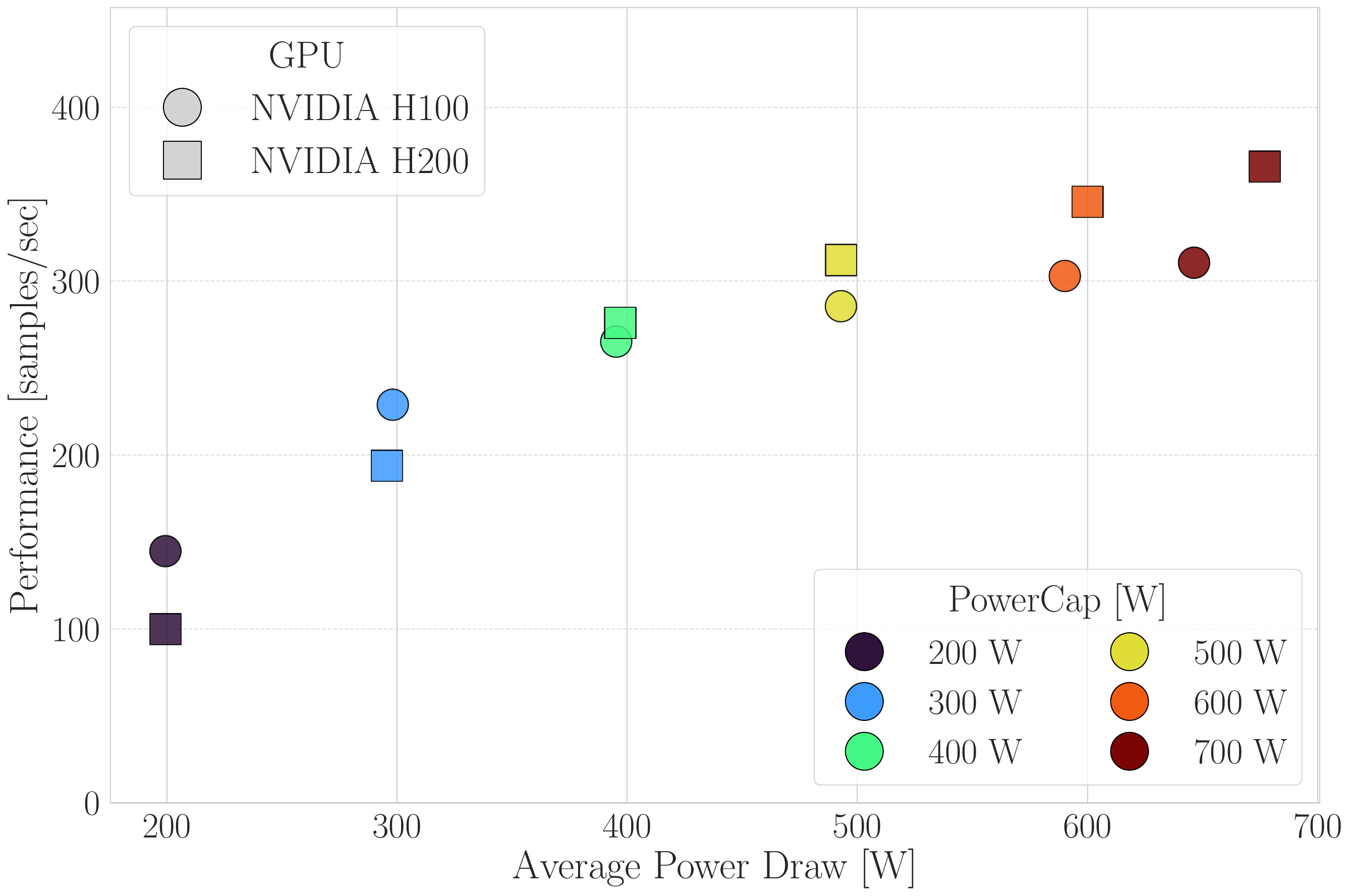}
        \caption{Performance throughput vs power draw}
        \label{fig:zplot_vit}
    \end{subfigure}
    \hfill 
    \begin{subfigure}[b]{0.49\textwidth}
        \includegraphics[width=\textwidth]{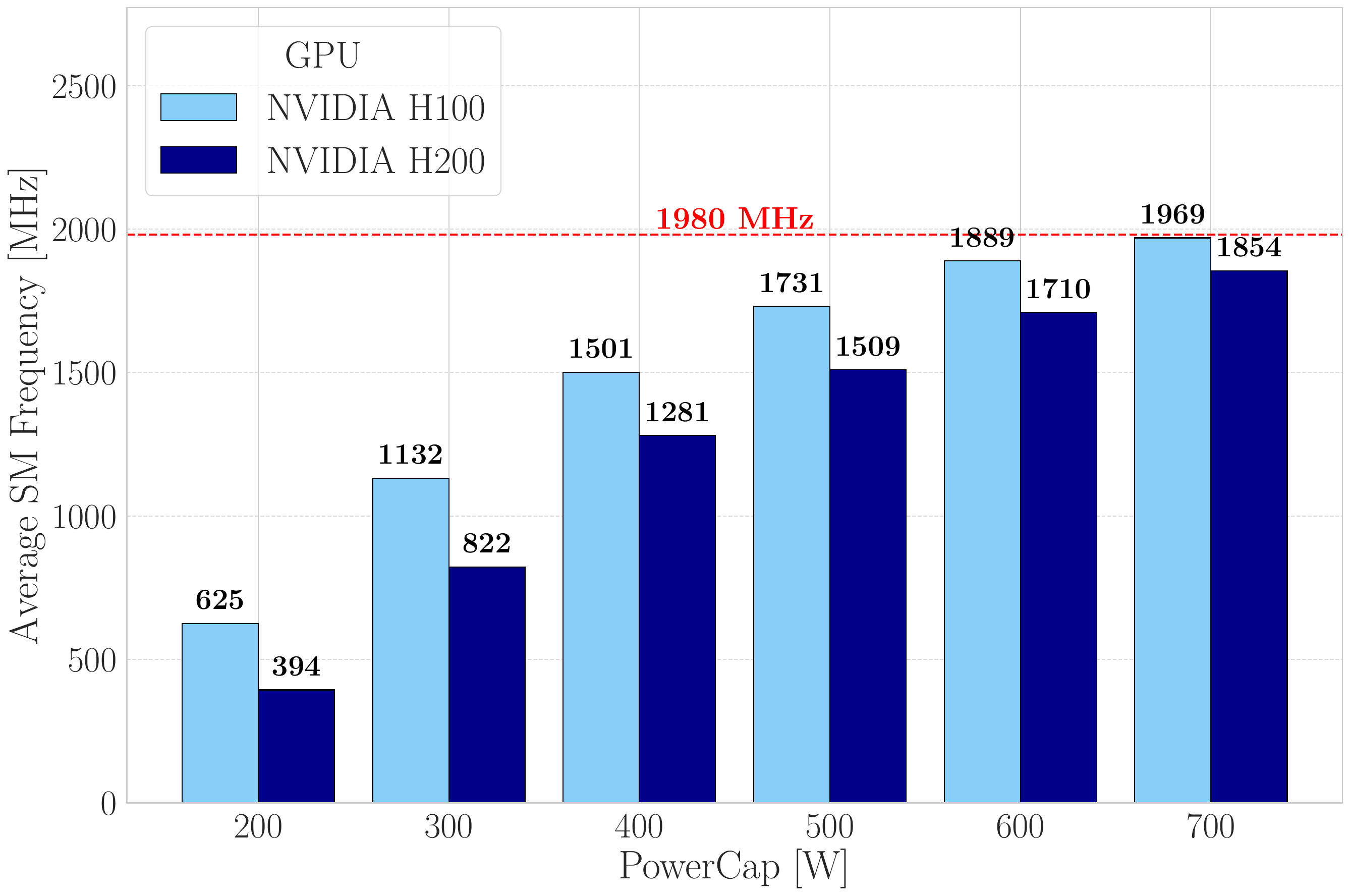}
        \caption{Average SM frequency vs different power-cap}
        \label{fig:vit_frequency}
    \end{subfigure}
    \caption{Relationship between different metrics for the ViT benchmark}
    \label{fig:zplot_vit}
\end{figure}

Figure~\ref{fig:zplot_vit} illustrates trends for the ViT workload. On the H100, the ViT benchmark does not quite hit the TDP, whereas on the H200, it triggers SM frequency throttling upon reaching TDP. Furthermore, at lower power caps for the H200, ViT exhibits intensified SM frequency throttling, almost dropping down to the base operating SM frequency of 345~MHz and causing a sharp decline in performance at 200~W. We observed the same before for STriad frequency throttling behavior in Figure~\ref{fig:sv_striad_freq}.
Interestingly, the H100 performs better at lower power caps and H200 performs better at higher power caps, with almost the same performance for both GPUs at 400~W. This is because ViT benefits from the higher bandwidth of H200, while it suffers from higher SM frequency throttling at lower power caps.

\section{Efficiency and Power Breakdown Comparison}
\label{sec:5}

This section evaluates GPU performance in terms of energy efficiency (performance per Watt). To contextualize the efficiency trends, we provide a comparative power breakdown, analyzing the SM and memory power draw for both the H100 and H200.

\subsection{DGEMM}

\begin{figure}[tbp]
    \centering
    \begin{subfigure}[b]{0.49\textwidth}
        \includegraphics[width=\textwidth]{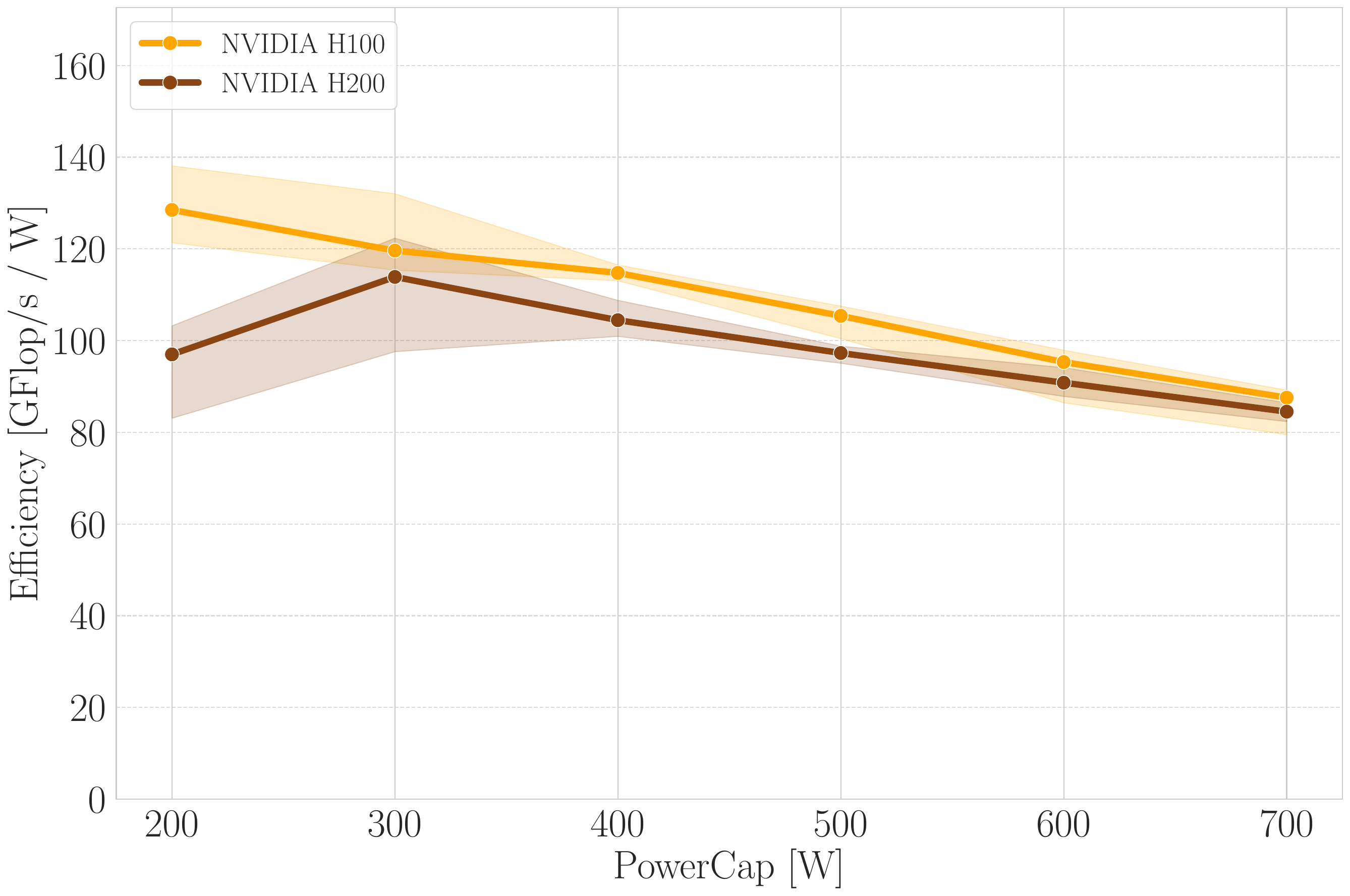}
        \caption{Efficiency Analysis}
        \label{fig:dgemm_efficiency}
    \end{subfigure}
    \hfill 
    \begin{subfigure}[b]{0.49\textwidth}
        \includegraphics[width=\textwidth]{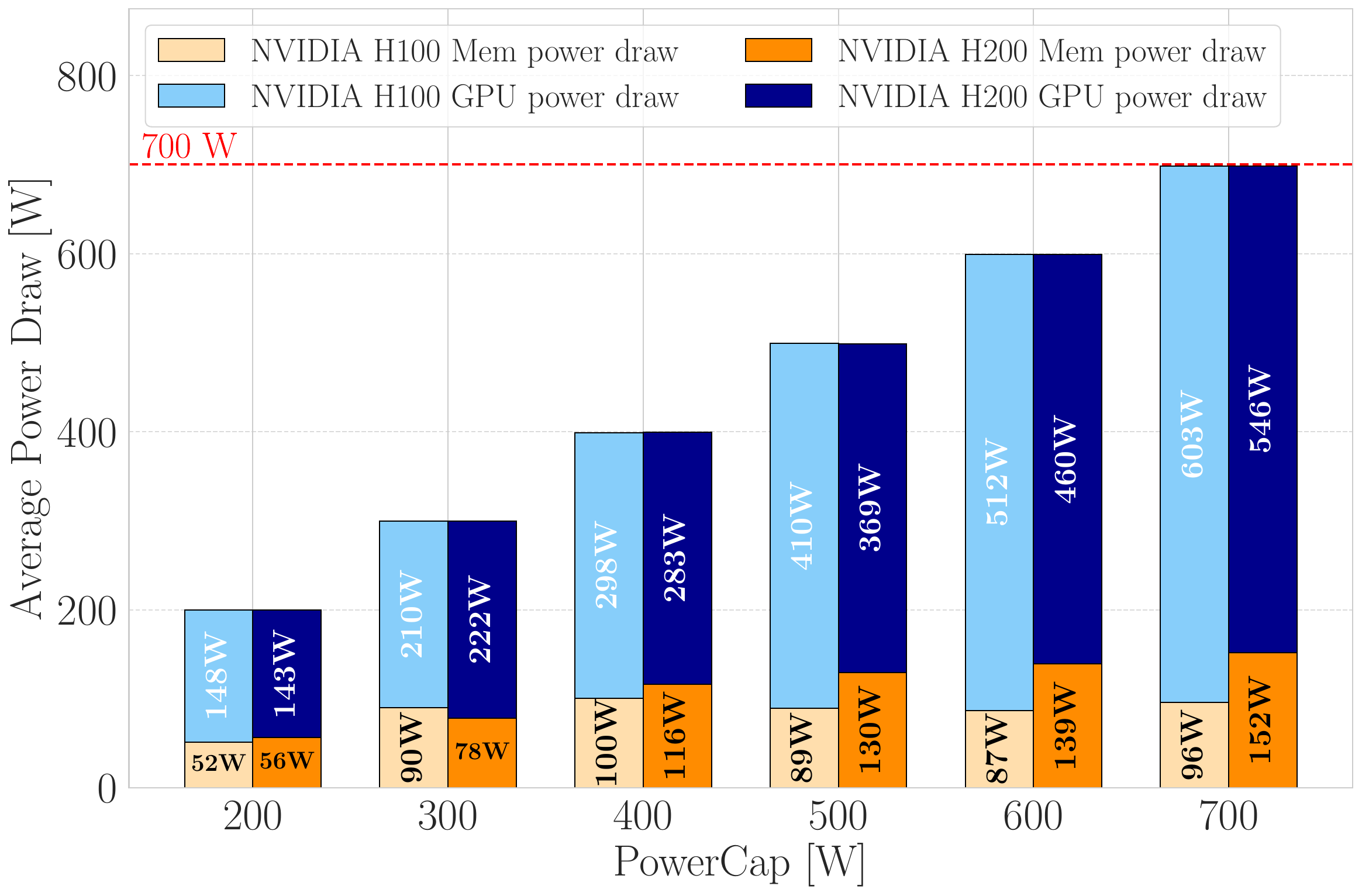}
        \caption{Power Breakdown}
        \label{fig:dgemm_power}
    \end{subfigure}
    \caption{DGEMM benchmark analysis: (a) Energy efficiency comparison between the NVIDIA H100 and H200, including min/max deviations, and (b) the corresponding power consumption breakdown between memory and GPU power.}
    \label{fig:dgemm_combined}
\end{figure}

Figure~\ref{fig:dgemm_efficiency} shows that both GPUs generally achieve higher energy efficiency under stricter power caps.
A notable exception occurs at the 200\,W limit for the H200, where disproportionately high memory power draw forces aggressive SM frequency throttling, severely degrading performance.
Additionally, both architectures exhibit a pronounced efficiency variance at power caps at or below 300\,W, driven by memory power fluctuations that destabilize GPU power draw and amplify SM frequency variability, as evident from Figure~\ref{fig:SV_DGEMM}. 

Figure~\ref{fig:dgemm_power} shows that the H200's memory power consumption continues to scale with the total power limit, correspondingly reducing the power budget available to the SMs.
Notably, despite a substantial gap in GPU power draw between the two GPUs up to the 500\,W cap, the resulting efficiency differences (Figure~\ref{fig:dgemm_efficiency}) remain marginal, suggesting that SMs can sustain near-optimal operating frequencies provided a sufficient baseline GPU power threshold is met.

\subsection{Schönauer Triad}
\label{STriad_efficiency}

\begin{figure}[tbp]
    \centering
    \begin{subfigure}[b]{0.49\textwidth}
        \includegraphics[width=\textwidth]{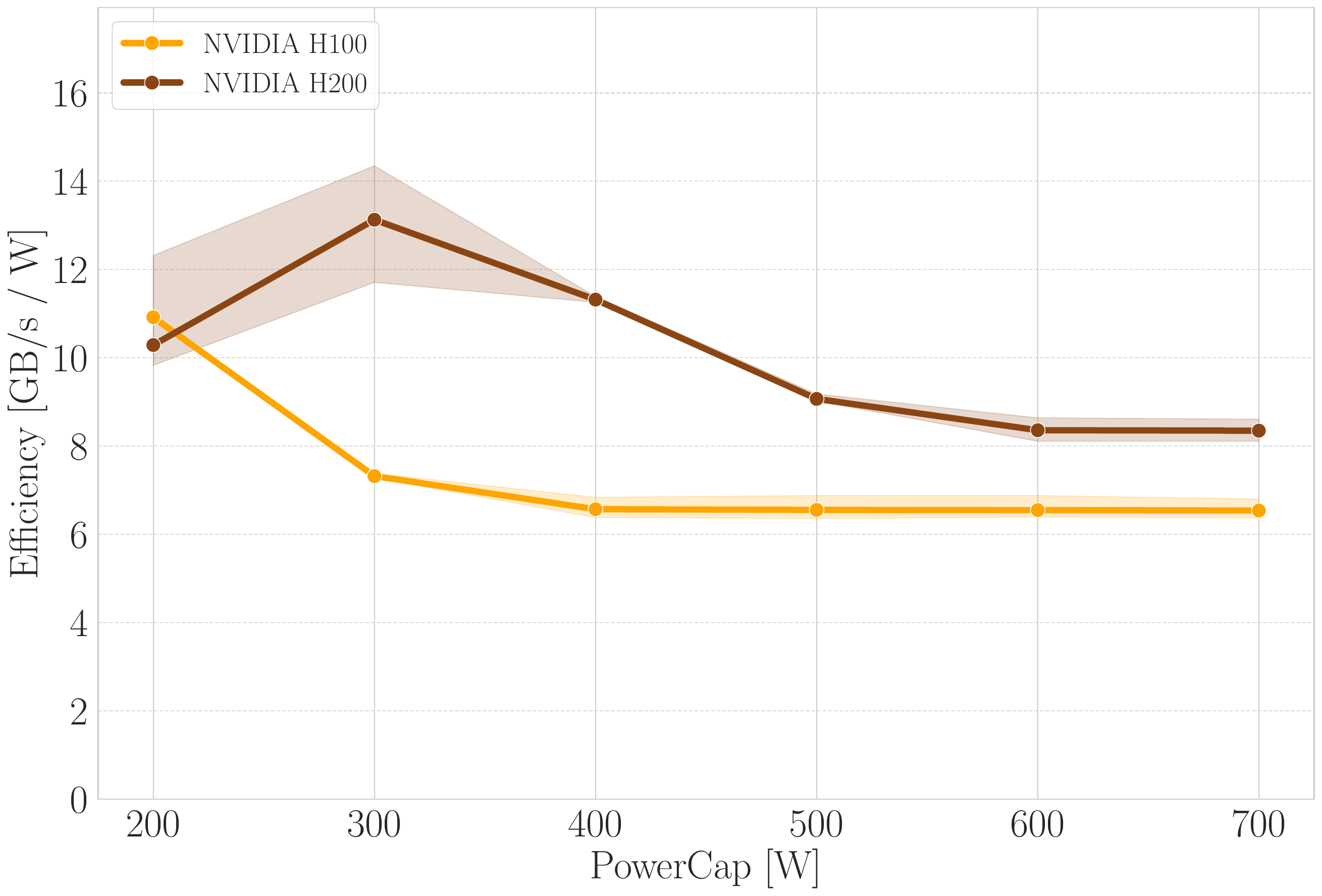}
        \caption{Efficiency Analysis}
        \label{fig:striad_efficiency}
    \end{subfigure}
    \hfill 
    \begin{subfigure}[b]{0.49\textwidth}
        \includegraphics[width=\textwidth]{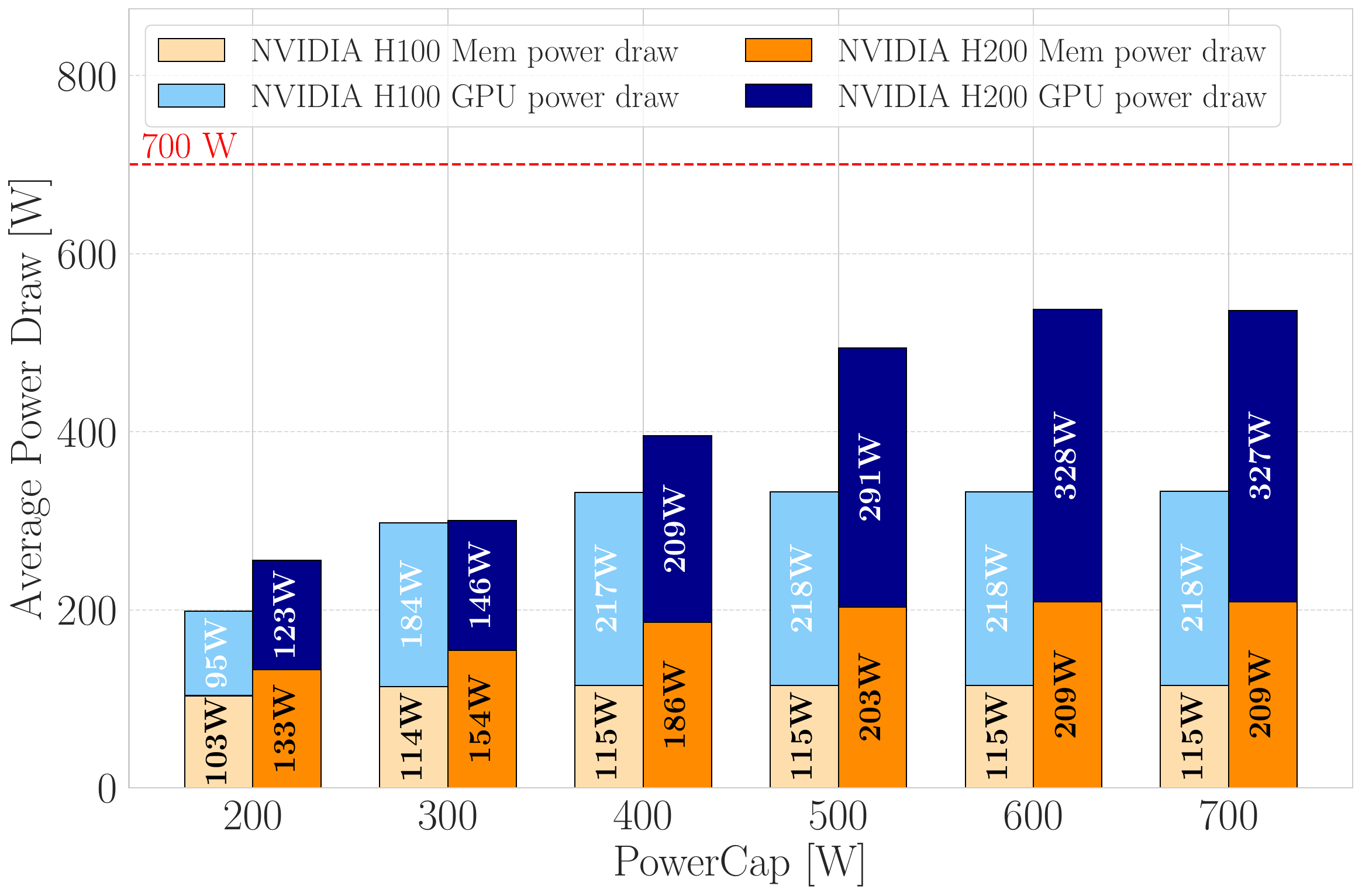}
        \caption{Power Breakdown}
        \label{fig:striad_power}
    \end{subfigure}
    \caption{Schönauer Triad benchmark analysis: (a) Energy efficiency comparison between the NVIDIA H100 and H200, including min/max deviations, and (b) the corresponding power consumption breakdown.}
    \label{fig:striad_combined}
\end{figure}

Figure~\ref{fig:striad_efficiency} reveals a clear energy efficiency advantage for the H200 across all power caps for the memory-bound benchmark. Notably, the H200 at a 400\,W limit outperforms the H100 at full TDP in overall efficiency. As observed previously, efficiency variance is pronounced at power caps at or below 300\,W, reflecting the underlying performance instability.

The comparative power breakdown in Figure~\ref{fig:striad_power} reveals that the H200 draws significantly more power across both its memory and SMs.
This increased SM power consumption is a direct consequence of the H200's superior memory bandwidth: The faster data delivery sustains a higher sustained flop/s throughput, thereby demanding more compute power.
Furthermore, the H200's doubled memory frequency relative to the H100 results in proportionally doubled memory power draw, provided total consumption remains within the power cap.

This power breakdown also explains the unusual 200\,W case where the H200 fails to stay under the power limit, unlike the H100.
Despite identical compute units, the H200's higher memory power draw significantly reduces the power budget available for computation.
The result can be seen in Figure~\ref{fig:sv_striad_freq}: The H200's SMs drop to their lowest supported SM frequency to stay within its power cap.
Still, performing more flop/s (due to higher memory bandwidth) at lowest operating SM frequency draws more power; in combination with memory power, it breaches the 200\,W power-cap setting.
We saw earlier in Figure~\ref{fig:dgemm_power} that the H200 can successfully stay under a 200\,W cap during compute-heavy workloads like DGEMM.
However, the massive memory power required by the memory-bound code makes it impossible for the H200 to obey the same 200\,W limit in this case.

\subsection{ViT}
\begin{figure}[tbp]
    \centering
    \begin{subfigure}[b]{0.49\textwidth}
        \includegraphics[width=\textwidth]{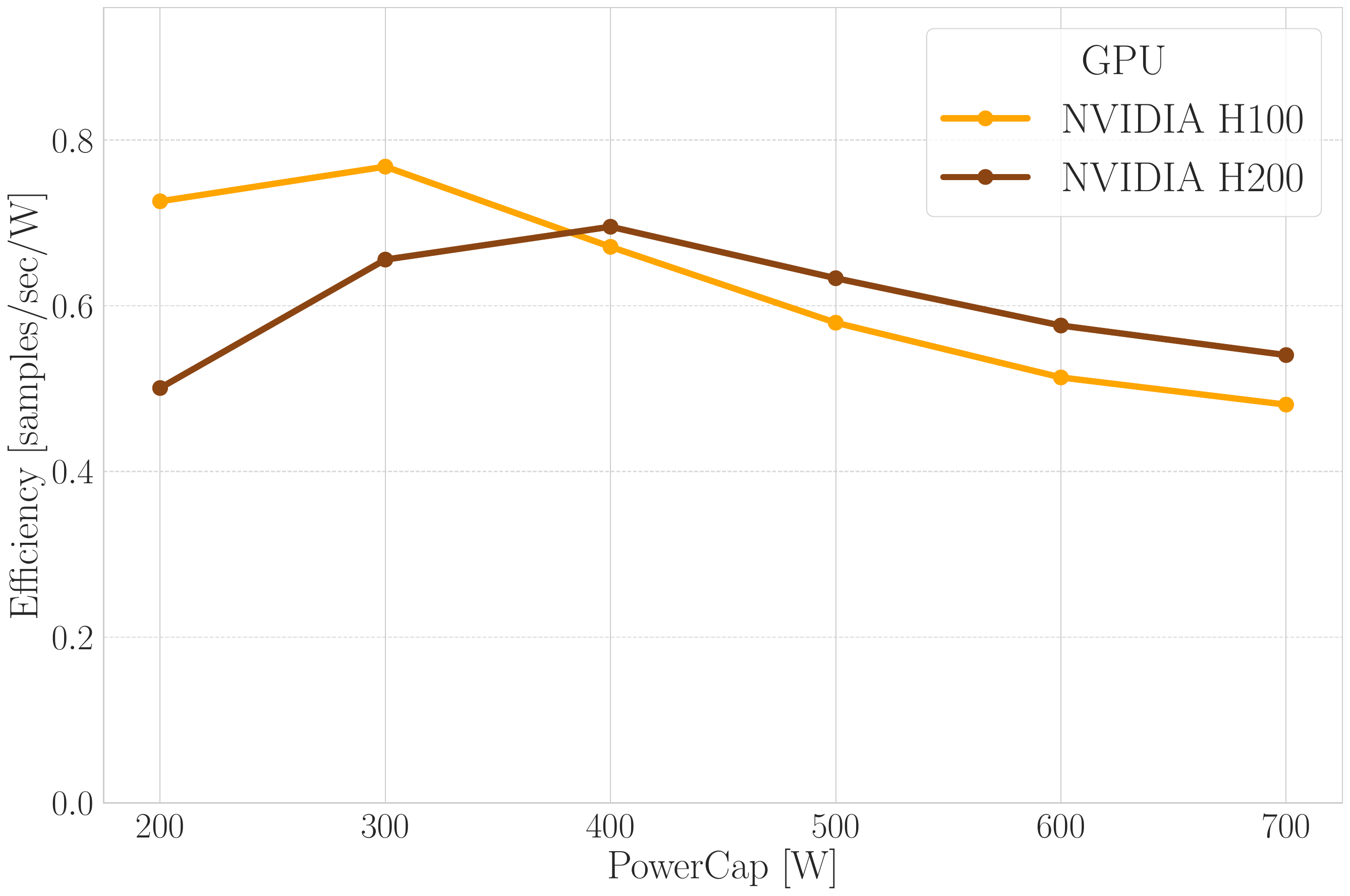}
        \caption{Efficiency Analysis}
        \label{fig:vit_efficiency}
    \end{subfigure}
    \hfill 
    \begin{subfigure}[b]{0.49\textwidth}
        \includegraphics[width=\textwidth]{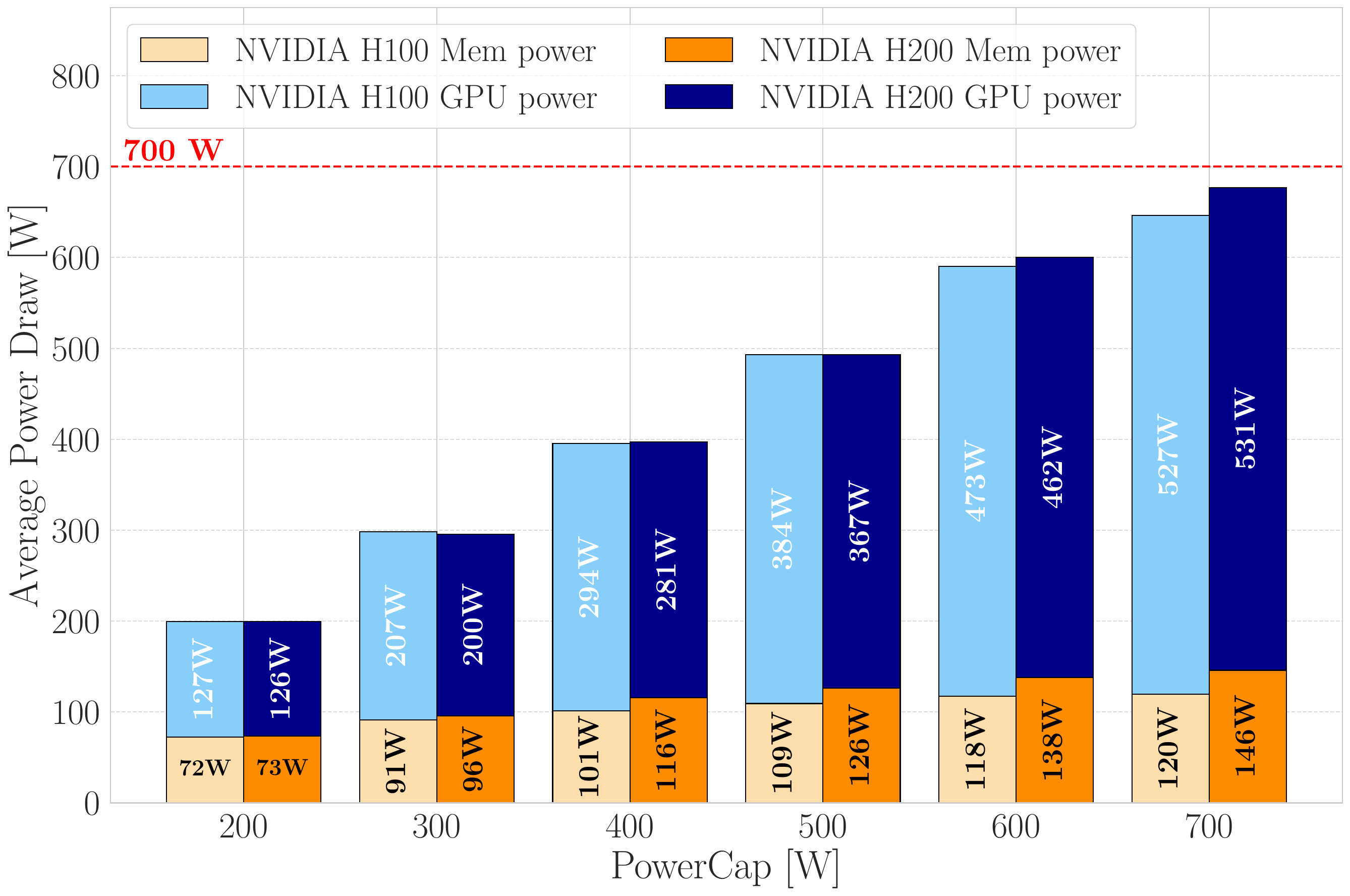}
        \caption{Power Breakdown}
        \label{fig:vit_power}
    \end{subfigure}
    \caption{ViT benchmark analysis: (a) Energy efficiency comparison between the NVIDIA H100 and H200, and (b) the corresponding power consumption breakdown between memory and GPU power.}
    \label{fig:vit_combined}
\end{figure}

The power-capping and power breakdown analysis for the Vision Transformer (ViT) benchmark shows a mix of trends compared to the micro-benchmark results. At higher power limits (400W or more), the ViT benchmark runs more efficiently on the H200 because it takes advantage of the H200's higher memory bandwidth. At lower power limits, however, the H100 is slightly more efficient. As shown in Figure~\ref{fig:vit_frequency}, the ViT benchmark causes more SM frequency throttling on the H200. Yet, Figure~\ref{fig:vit_power} shows that both GPUs draw almost the exact same amount of memory and total GPU power.

 When the power cap is raised, there is plenty of headroom to maintain a high SM operating frequency while still getting the full performance boost from the higher memory bandwidth, as seen in Figure~\ref{fig:zplot_striad_h200}.

\section{Conclusion}
\label{sec:conclusion}

Our analysis shows that the NVIDIA H100 slightly outperforms the H200 in compute-bound tasks (DGEMM) at the same power limits.
This is because the H200 requires more power for its memory, which then leaves less room for GPU power draw, resulting in lower operating SM frequency.
We also see that the performance for both GPUs is directly proportional to the SM frequency. The performance will scale according to the SM frequency scaling.
For both GPUs, increasing the power cap from 200\,W to 400\,W provides the best performance boost, while anything above 500\,W offers diminishing returns.

Conversely, the H200's superior memory bandwidth more than compensates for its higher power draw on memory-bound tasks, yielding substantially better energy efficiency.
Operating the H200 at a restricted 400\,W limit surpasses the H100 at full TDP in efficiency, though reaching peak memory bandwidth still requires 550\,W versus the H100's 350\,W.
However, higher memory bandwidth does not automatically guarantee better energy efficiency. It significantly increases the power consumption of both the memory and the SMs.
Most notably, the H200 fails to respect a strict 200\,W cap under memory-intensive loads, drawing approximately 250\,W instead: Its high baseline memory power consumption leaves insufficient headroom for the SMs, forcing them to their lowest supported SM frequency.

Comparing micro-benchmark and AI workload results demonstrates that micro-benchmarks are highly effective for uncovering hardware extremes, providing insights that directly translate to real-world applications. Crucially, a workload's efficiency at lower power caps depends on its arithmetic intensity within the Roof\-line model. The application's arithmetic intensity determines its memory power draw, which consequently dictates performance by forcing the SM frequency to throttle based on the remaining power budget.

Additionally, our regression testing successfully uncovered undocumented hardware properties (like memory power draw) and highlighted distinct hardware outliers, proving that individual GPUs of the exact same model can consistently draw measurably more memory power than other GPUs of the same type.
We observe maximum memory power draw of 120\,W for H100 and maximum memory power draw of 220\,W for H200 (except the outliers at 240\,W).

\begin{credits}
\subsubsection{\ackname} The authors gratefully acknowledge the scientific support and HPC resources provided by the Erlangen National High Performance Computing Center (NHR@FAU) of the Friedrich-Alexander-Universität Erlangen-Nürnberg (FAU) under the BayernKI project v111dc. BayernKI funding is provided by Bavarian state authorities.
The authors gratefully acknowledge the scientific support and HPC resources provided by NHR@FAU of the FAU under the NHR project b104dc. NHR funding is provided by federal and Bavarian state authorities. NHR@FAU hardware is partially funded by the German Research Foundation (DFG) -- 440719683.

\end{credits}
%
%
%
\bibliography{ref}

\bibliographystyle{splncs04}

\end{document}